\documentclass[preprint,aps,eqsecnum,nofootinbib]{revtex4}

\usepackage{epsfig}

\newcommand{\be}{\begin{equation}}
\newcommand{\ee}{\end{equation}}
\newcommand{\bea}{\begin{eqnarray}}
\newcommand{\beas}{\begin{eqnarray*}}
\newcommand{\eea}{\end{eqnarray}}
\newcommand{\eeas}{\end{eqnarray*}}
\newcommand{\ba}{\begin{array}}
\newcommand{\ea}{\end{array}}

\newcommand{\la}{\langle}
\newcommand{\ra}{\rangle}
\def\ls{\mathrel{\lower4pt\vbox{\lineskip=0pt\baselineskip=0pt
           \hbox{$<$}\hbox{$\sim$}}}}
\def\gs{\mathrel{\lower4pt\vbox{\lineskip=0pt\baselineskip=0pt
           \hbox{$>$}\hbox{$\sim$}}}}

\begin{document}


\title{Fixing the Solar Neutrino Parameters with Sterile Neutrinos}
\author{J.C. G\'omez-Izquierdo\footnote{e-mail:jcarlos@fis.cinvestav.mx}
and A. P\'erez-Lorenzana\footnote{e-mail:aplorenz@fis.cinvestav.mx} }

\affiliation{
Departamento de F\'{\i}sica, Cinvestav. 
Apdo. Post. 14-740, 07000, M\'exico, D.F., M\'exico}

\date{May, 2006}

\begin{abstract}

Neutrino mixing matrix appears to be close to bimaximal mixing, but for the
solar mixing angle which is definitively smaller than forty five degrees.  
Whereas it seems  quite easy to understand bimaximal mixing  with the use of
new global symmetries, as in models using $L_e - L_\mu - L_\tau$,  
understanding the about to 
eleven degrees of deviation in the observed solar angle
seems less simple. We suggest that such a deviation could be due to a light
sterile neutrino that mixes with the active sector. 
The mass scale needed to
produce the effect has to be  smaller than atmospheric scale, and  it would 
introduce a new mass squared difference  which should be 
smaller than the solar scale.
We present a toy model that exemplifies these features.

\end{abstract}

\maketitle

\section{Introduction}
Convincing evidence that neutrinos have mass and oscillate has been
provided along recent years by  
Kamiokande, Super-Kamiokande, MACRO and Soudan 
results on atmospheric neutrinos; 
by Chlorine, Kamiokande,
Super-Kamiokande, SAGE, GALLEX and most recently the SNO experiment on solar
neutrinos; 
as well as by KamLAND, K2K and CHOOZ-PALO-Verde,  
base-line neutrino experiments~\cite{fogli}. 
KAMLAND independent  confirmation of solar oscillation parameters
observed by SNO data, 
indicates that the observed solar mixing is due to a large
mixing angle oscillations enhanced by the MSW matter effect~\cite{msw}. 

In the standard framework,  only three weak 
neutrino species,  $\nu_e$; $\nu_\mu$ and $\nu_\tau$, are needed 
to consistently  describe the mentioned experimental
results, with the only addition  of neutrino masses and mixings as  new
parameters to the Standard Model. Central idea in the oscillation phenomena is
that, as it happens in the quark sector, neutrino mass eigenstates,
$\nu_{1,2,3}$,  and weak eigenstates are different, 
but they can be written as linear combinations of
each other by using a complex unitary matrix, $U$, as
$\nu_{\alpha } = \sum_i U_{\alpha i}\nu_{i}$, for $\alpha=e,\mu,\tau$ and
$i=1,2,3$, where we refer only to left handed states.
A common parameterization for Majorana neutrinos 
of the $U$ matrix is given in terms of three angles
and three CP phases, such that  $U = U_{PMNS} K$,  where  
$K={\rm diag}\{ 1,e^{i\phi_1},e^{i\phi_2}\}$, with $\phi_1$, $\phi_2$ the
physical CP-odd Majorana phases. The elements of the
Pontecorvo-Maki-Nakagawa-Sakata (PMNS) matrix~\cite{pmns} are then 
 \be 
 U_{PMNS} = \left(\ba{ccc} 
 c_{12} c_{13} & s_{12}c_{13} &  s_{13} e^{-i\varphi} \\
 -s_{12}c_{23} - c_{12}s_{23}s_{13}e^{i\varphi} &
 c_{12}c_{23} - s_{12}s_{23}s_{13}e^{i\varphi} & s_{23}c_{13} \\
 s_{12}s_{23} - c_{12}c_{23}s_{13}e^{i\varphi} &
 -c_{12}s_{23} - s_{12}c_{23}s_{13}e^{i\varphi} & c_{23}c_{13} \ea\right)~;
 \ee
where $c_{ij}$ and $s_{ij}$ stand for $\cos\theta_{ij}$ and $\sin\theta_{ij}$
respectively. They represent the observable mixing angles in the basis where the
charged lepton masses are diagonal.
The  Dirac CP phase, $\varphi$, is the only phase involved in
neutrino oscillations. Most analysis of neutrino data are usually done
in the hypothesis that $\varphi$ is negligible.  This
is particularly correct in the case for solar and reactor oscillations data. 
We will assume so for simplicity hereafter.
Finally the kinematical  scales for the oscillation 
are  given by the two mass squared
differences: (i) the solar/KamLAND scale $\Delta m^2_{sol}= \Delta m_{12}^2$;
and (ii) the atmospheric scale $\Delta m^2_{ATM}= \Delta m_{23}^2$. 

Combined analysis of all data indicates that at two sigma level~\cite{fogli}
 \bea 
 \Delta m^2_{sol} &=& (7.92 \pm 0.71) \times 10^{-5}~{\rm eV^2}~; \nonumber\\
 \Delta m^2_{ATM} &=& (2.4~^{+0.5}_{- 0.62}) \times 10^{-3}~{\rm eV^2}~;
 \eea
for the absolute scales, and 
 \be 
 \sin^2\theta_{12}  = 0.314~^{+0.057}_{-0.047}~; \quad
 \sin^2\theta_{23}  = 0.44~^{+0.18}_{-0.096}~.
 \ee
CHOOZ-Palo Verde data provide the stringent constraint on $\theta_{13}$.
Again, the analysis in Ref.~\cite{fogli} gives
 \be 
 \sin^2\theta_{13}  < 0.9~(+0.207)\times 10^{-2}~;
 \ee
where the number within parenthesis stands for the two sigmas upper uncertainty.
It is particularly
interesting to notice that two of the above mixing angles are rather large. 
Also, the fact that $\theta_{13}$ is consistent with zero 
indicates that solar/KamLAND experiments are mainly sensible to $\theta_{12}$,
which means that $\sin^2\theta_{sol}= \sin^2\theta_{12}$; 
whereas the mixing observed in atmospheric neutrinos is basically given by 
$\sin^2\theta_{ATM}= \sin^2\theta_{23}$. Thus, atmospheric muon neutrino
deficit  is due to
maximal (or almost maximal) mixing among muon and tau neutrinos; whereas solar
deficit is due to a large, but not maximal, mixing of electron to other
active neutrino species. Indeed, by
taking  central values of the mixing angles one  sees that
 \be
 \theta_{12} \approx 34.08^o~; \quad \mbox{and}\quad\theta_{23} \approx 41.55^o~.
 \ee
Thus, solar mixing is far from being maximal by at least eleven degrees.

A complete understanding of the   value of neutrino oscillation
parameters  from a theoretical
point of view is yet more challenging. On one hand side, it is tempting to 
belive that $U_{PMNS}$ could be the result of a relatively small perturbation
around the bimaximal mixing matrix
 \be
 U_{BM} = 
 \left(\ba{ccc}
 \frac{1}{\sqrt{2}} & \frac{1}{\sqrt{2}} & 0\\[1ex]
 \frac{1}{2} & -\frac{1}{2} & \frac{1}{\sqrt{2}} \\[1ex]
 \frac{1}{2} &  -\frac{1}{2} & -\frac{1}{\sqrt{2}}
 \ea \right)~; 
 \label{ubm1}
 \ee
for which two angles are exactly maximal and the third is null. 
This at least seems to be  a very good approximation 
for atmospheric and $\theta_{13}$ angles within one sigma level. 
One could then assume that this matrix
arises as the zero order of a theory for neutrino 
masses and mixings that contains some 
global (flavor) symmetry $G$, which is actually broken in a way that the
amount of breaking of $G$ would provide the eleven degrees of deviation in
solar angle. One would then write 
\be 
U_{PMNS} = U_{BM}\cdot U_{A}~,
\ee
where $U_A$ parameterizes the additional rotations induced by the breaking of
the flavor symmetry. Indeed there are simple models already in the literature
that realize bimaximal mixing. This happens for instance in models  that use 
$L' =  L_e -L_\mu - L_\tau$ as a global symmetry for the neutrino mass
matrix~\cite{numodels,pires}.
On the other hand, it is a typical feature of those models to induce the desired
correction in the neutrino mixings via loop effects, which are usually
suppressed to the level of being smaller than 
what is required for an  understanding of the solar mixing. 

Recently it has also  been suggested  that $U_{A}$ in last equation 
could  be  the very  same CKM matrix $U_{CKM}$
of the quark sector~\cite{qlcomp}. This quark-lepton complementarity 
is indeed an intriguing possibility since  the Cabbibo angle is just about
twelve degrees.  Nevertheless, its realization seems to require 
that quark and lepton masses be somehow correlated in a non
trivial way, and so far there are no complete models that may satisfactorily 
realize it~\cite{qlcmodels}. 

In this paper we will take a different perspective to the problem, and 
suggest that the $U_A$ correction may  rather come due 
to the couplings of the active neutrinos to a  
fourth (sterile) neutrino which is lighter than the atmospheric scale, and thus
it is not constrained by the LSND nor Bugey/CHOOZ nor KARMEN
data. 
Our hypothesis may rather be constrained by solar data, 
however,  with the currently  allowed range of active-sterile mixing at one
sigma level, $\sin^2\eta<0.09$~\cite{vissani},  there may still be  enough room as to provide
the desired corrections.  
We should mention that  
light sterile neutrinos were suggested earlier in
Ref.~\cite{holanda} as a way to fix a small deviation of Homestake Ar
production rate results from the generic LMA prediction,  and the apparent
absence of the upturn  of the energy spectrum at low energies in
Super-Kamiokande and SNO. We don not analyze this possible effects here,
since our main goal for the moment 
is to present an additional 
possible  theoretical use of the light sterile neutrinos. 
To be more specific in the discussion we
will analyze the particular case of models for inverted hierarchy that
use a global $L'=L_e -L_\mu -L_\tau$ symmetry.

The paper is organized as follows. In section 2 we discuss the generalities of
the  $L'$ models that provide bimaximal mixing. For
completeness we show that the squared mass spectrum is inverted, with the two
heavier states degenerated at the limit of the exact $L'$ symmetry. We then show
that a generic diagonal correction on the mass matrix,  which explicitely breaks
the $L'$symmetry,  does not provide enough freedom to simultaneously generate
the solar mass splitting and the eleven degrees of corrections to the solar
mixing. Either one of them comes out to be larger than the observed values.   In
section 3 we address the question of whether the breaking of the $L'$ symmetry
by the coupling with a sterile neutrino may do the job.  As the sterile is
expected to alter the solar parameters with out substantially affecting the
atmospheric ones, we will work in the hypothesis that the sterile mixes
preferentially to a single active state after a bimaximal rotation. 
Thus,  we 
show that within this hypothesis the  sterile couplings by themselves are
also unlikely to provide both the solar parameters, however,  
both the effects, the $L'$ diagonal 
breaking mass terms on the active sector and the sterile 
coupling, may compensate each other 
to provide the solar mass scale and, at the same time, to give  the right
solar mixing. We present the results of a numerical analysis
to identify the narrow 
region on the parameter space where our mechanism may work for the particular
texture  that we introduce. Finally, 
some concluding remarks are presented.

\section{The Solar Mixing Problem in $L_e -L_\mu -L_\tau$ Models.}

We start by assuming that
the Majorana neutrino mass matrix in the basis $( \nu_e, \nu_\mu, \nu_\tau)$
has the form 
 \be 
 M_0 = m \left(\ba{ccc} 
 0 & \cos\theta & \sin\theta \\
 \cos\theta & 0 & 0\\
 \sin\theta & 0 & 0 \\
 \ea\right)~;
 \label{mo}
 \ee
where the overall scale sets the atmospheric scale,
$m=\sqrt{\Delta m_{ATM}^2}$, and $\theta$ shall correspond to 
the atmospheric angle, which for the present analysis will be 
treated as general although in our final estimations it would be taken 
to be exactly $\pi/4$. 
There are many models nowadays in the literature that 
provide the above given mass texture, see for instance
references~\cite{numodels,pires}.  
An attractive way to motivate such mass terms is by
assuming that the tree level Majorana mass terms obey a global 
$L' =  L_e -L_\mu - L_\tau$. Consider for instance a simple model 
with  total lepton number and $L'$ as global symmetries, 
where, besides the Standard Higgs doublet $H(L=0,L'=0)$, there is 
a scalar triplet $\Delta(L=2,L'=0)$. 
Since our triplet has no $L'$ charge, the only allowed Yukawa couplings
involved in  neutrino masses are
 \[
 \Delta~\bar L_e^c(a L_\mu + b L_\tau) + h.c~;
 \]
where $a$ and $b$ stand for the corresponding Yukawa couplings. 
Note that we choose to work  in the basis where Dirac  Yukawa couplings
and thus charged lepton masses are diagonal for this
field content. 
After symmetry
breaking, assuming that $\la \Delta\ra\neq 0$ and small, one gets the Type II
seesaw neutrino masses that are given in the mass matrix of Eq.~(\ref{mo}), 
with  $m = \la\Delta\ra\sqrt{a^2 + b^2}$ and $\tan\theta = a/b$.

It is easy to see that the more general form of the bimaximal mixing matrix 
 \be 
 U_{BM} = 
 \left(\ba{ccc}
 \frac{1}{\sqrt{2}} & \frac{1}{\sqrt{2}} & 0\\[1ex]
 \frac{\cos\theta}{\sqrt{2}} & -\frac{\cos\theta}{\sqrt{2}} & \sin\theta \\[1ex]
 \frac{\sin\theta}{\sqrt{2}} &  -\frac{\sin\theta}{\sqrt{2}} & -\cos\theta
 \ea \right)~; 
 \label{ubm}
\ee
which for $\theta=\pi/4$ reduces to that given in  Eq.~(\ref{ubm1}); 
diagonalizes $M_0$. Indeed, 
\be
U_{BM}^\dagger M_0 U_{BM} = 
m\left(\ba{ccc} 1 & 0 & 0 \\  0 &-1&0 \\ 0 &0&0 \ea\right)~.
\ee
Thus, the spectrum is inverted and the heavier squared masses are degenerated.
The gap $\Delta m^2_{13} = \Delta m^2_{23} = m^2 $ 
is identified with the atmospheric scale as already mentioned, and thus,
 $\theta$ becomes the atmospheric mixing angle. 
At this level, the model does not provide any explanation for the solar
scale, since $\Delta m^2_{12}$ is exactly zero. Moreover, the mixing angle 
calculated from the standard 
formula for the solar mixing:  $\sin^2 2\theta_{sol} = 4 (U_{e1}U_{e2})^2$; 
exactly gives $\theta_{sol} = \pi/4$. Of course, strictly speaking this simple
model predicts no solar neutrino oscillations, and the mentioned angle is only
for reference  proposes.  

The problem is relieved when one realizes that 
the zeros that appeared on the $M_0$
texture are actually representing some 
small numbers, which would  be introduced by
the global symmetry breaking effects that we have neglected so far.
In many models this small terms may come from radiative
corrections~\cite{numodels} or even from non renormalizable
operators~\cite{pires}.
To be specific let us consider the model presented in Ref.~\cite{s3model}. 
Apart from the standard lepton content, 
we introduce three Higgs doublets $\phi_{0,1,2}$, two triplets,
$\Delta_{1,2}(Y=2)$ and one isosinglet, $\eta^+(Y=2)$. We impose the symmetry
$L'\times S_3$, where $S_3$ is the permutation group of three elements, under
which $2_L=(L_\mu,L_\tau)$, $2_R=(\mu_R, \tau_R)$, $2_\Phi=(\phi_1,\phi_2)$ and 
$2_\Delta=(\Delta_1,\Delta_2)$ are doublets, 
with all other fields as $S_3$ singlets, but with 
$\eta$ an odd (pseudo) singlet. 
The allowed Yukawa couplings can be written in compact notation as 
\[ {\cal L}_Y = h_1~\bar{2}_L\cdot 2_R\phi_0~ + 
h_2~\bar{2}_L\times 2_R \cdot 2_\Phi + h_e \bar{L}_e e_R\phi_0 + 
fL_e 2_L\cdot 2_\Delta + f' L_\mu L_\tau\eta + h.c~;
\]
where the indicated doublet products represent the 
$S_3$ invariants obtained as follows: given the
$S_3$ doublets $2_x = (x_1,x_2)$ and $2_y = (y_1,y_2)$, we  built the
even singlet $1_{xy}= 2_x\cdot 2_y = x_1 y_1 + x_2 y_2$ 
that gives the first and
fourth terms in ${\cal L}_Y$; the odd singlet 
$1'_{xy} = x_1 y_2 - x_2 y_1$ that gives the coupling to $\eta$ field, 
and the new doublet 
$2_{xy} =2_x\times 2_y = (x_1 y_1 - x_2 y_2 , x_1 y_2 + x_2 y_1)$
involved in the second Yukawa term above. 

To one loop order, one gets the
neutrino mass matrix
 \be 
 M_1 = M_0 + M_\epsilon = m \left(\ba{ccc} 
 0  & \cos\theta & \sin\theta \\
 \cos\theta & \epsilon & 0\\
 \sin\theta & 0 & \epsilon \\
 \ea\right)~.
 \label{m1}
 \ee
where the diagonal terms come from a one loop graph as the one shown in 
Fig.~1. They are about same order, hence we assume them equal.
The model gives no charged lepton mixing at tree level,
thus, from the diagram is easy to see why there are 
no off-diagonal mass terms  generated at one loop. Also,  $\eta$
does not couple to $L_e$, which explains why  $m_{ee}=0$ at the same order. 
Hereafter, for our analysis we will only required to assume 
the above texture, in the understanding that 
other  models could certainly fulfill the same neutrino mass structure.  
 Notice however, that this is certainly not the
most general texture, but it serves very well to our propose of 
motivating the possible sterile corrections.

\begin{figure}
\centerline{
\epsfxsize=200pt
\epsfbox{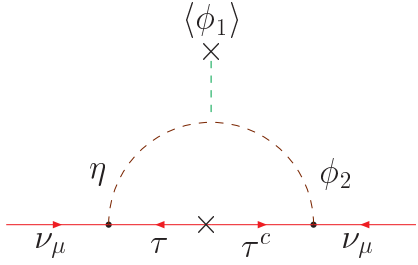}
}
\vskip1ex

\caption{One loop diagram that generates the $m_{\mu\mu}$ term. A similar
diagram would provide $m_{\tau\tau}$.}

\end{figure}

After introducing the bimaximal rotation, one may write the  perturbation as 
 \be 
 U_{BM}^\dagger M_\epsilon U_{BM} = 
 \epsilon~ m \left(\ba{rrc} \frac{1}{2} & -\frac{1}{2} & 0 \\  
 -\frac{1}{2} &\frac{1}{2}&0 \\ 0 &0&1 \ea\right)~.
 \label{me}
 \ee
This expression already shows what it is desired from the $L'$
breaking effects: they should mainly modify $\theta_{12}$, and split the squared
masses of the heavier states whereas leaving almost unaffected the other
two mixing angles $\theta_{23}$ and $\theta_{13}$. 
As a plus, the lighter neutrino  gets a small 
mass $m_3 = \epsilon m$. However, since this mass is not constrained by
observations it can not be used to fix the new parameter. Clearly, only a new
rotation on the 1-2 plane
is needed to compensate for the effect of the
perturbation. An exact calculation shows that the rotation angle satisfies
 \[\tan\alpha = 
 \frac{\epsilon/2}
 {1 +\sqrt{1+\epsilon^2/4}}
 \approx \frac{\epsilon}{4}\]
Thus, the final mixing matrix  would be
 \be 
 U_{mix} = U_{BM}\cdot U_\alpha = \left(\ba{ccc} 
  \frac{c_\alpha-s_\alpha}{\sqrt{2}} & \frac{c_\alpha+s_\alpha}{\sqrt{2}} &0 \\
  \frac{c_\alpha+s_\alpha}{\sqrt{2}}c_\theta &
  \frac{s_\alpha-c_\alpha}{\sqrt{2}}c_\theta & s_\theta\\
  \frac{c_\alpha+s_\alpha}{\sqrt{2}}s_\theta& 
   \frac{s_\alpha-c_\alpha}{\sqrt{2}}s_\theta& -c_\theta\ea \right)
 \ee
where, as before, we have written $c_{\alpha}$ ($s_\alpha$) for 
$\cos\alpha$  ($\sin\alpha$) and $c_{\theta}$ ($s_\theta$) for 
$\cos\theta$  ($\sin\theta$) to
simplify the expression.
Therefore, the solar mixing angle  would now be given by 
 \be 
 \sin^22\theta_{sol} = (\cos^2\alpha - \sin^2\alpha)^2 = \cos^22\alpha~; 
 \label{solmixa} 
 \ee
or equivalently
 \be 
 \sin^22\theta_{sol}
 = \left[\frac{1+\sqrt{1+\epsilon^2/4}}
 {1 + \epsilon^2/4+\sqrt{1+\epsilon^2/4} }\right]^2
 \approx  1 - \frac{\epsilon^2}{4}~.
 \label{solmixa1} 
 \ee
It is interesting to note that, since $\cos2\alpha = \sin (\pi/2 - 2\alpha)$,
 equation (\ref{solmixa}) means that 
\[ \theta_{sol} = \frac{\pi}{4} - \alpha~.\]
This explicitely shows that the rotation introduced due to the 
$\epsilon$ corrections work in the right direction, by reducing the value of
the solar angle down from  maximal. Next question would be whether such
a correction would be enough to provide the right observed values. To answer
this question one has to consider the heavier mass eigenvalues:
$m_1 = m\left( \epsilon/2 + \sqrt{1+\epsilon^2/4} \right)$; and 
$m_2 = m \left( \epsilon/2 -\sqrt{1+\epsilon^2/4} \right)$;
from which, without approximations, one gets
 \be 
 \Delta m^2_{sol} = \epsilon~m^2 ~\sqrt{4+ \epsilon^2}~.
 \label{dma}
 \ee
Thus, 
 \be  
 \epsilon \approx \frac{1}{2}\frac{\Delta m^2_{sol}}{\Delta m^2_{ATM}}~;
 \label{epsilon0}
 \ee
which indicates that $\epsilon$, 
and  thus $\alpha$, are of the order of one
part in a hundred at most, and so they are rather much 
smaller than what is needed to give the solar mixing, according to
Eq.~(\ref{solmixa1}). 
Indeed, such a small $\epsilon$ would mean a correction on
$\sin^2 2\theta_{sol}$ on  just about one part on ten thousands. 
Using central values for the solar and atmospheric scales, and the exact
expressions (\ref{solmixa1}) and (\ref{dma})
given above, by fixing $\epsilon$ with the solar scale, one gets 
$\sin^22\theta_{sol} = 0.9999$, that means  $\sin^2\theta_{sol}= 0.496$ 
which is far larger than what it is actually  desired.
Conversely, if one takes $\epsilon$ large enough as to  fix the solar
mixing to the appropriate level, one gets a too large $\Delta m^2_{12}$.

\section{Solar Mixing and  the Sterile Neutrino}

From the analysis of previous section one can see that 
the simple correction we  considered is unable to 
provide  both the solar mass scale and mixing angle. 
A complete understanding of these parameters,  
along this line of thought, needs  to consider a more
complicated correction of the original texture, which may be an indication 
of a  quite complicated  way in which the associated $L'$ 
symmetry is been  broken~\cite{leontaris}. 
Another possibility is that the 
desired corrections may come from  
some other source. This external source could be the coupling with a fourth
neutrino. Such couplings usually violate the global $L'$ symmetry and
could be as large as the tree level masses without any fundamental
contradiction. Another question is, of course, the reason why a singlet field,
as a fourth neutrino,
comes with mass couplings as light as those of
the active sector, since they are not protected by any Standard Model symmetry. 
We will not address this question in here, nor give a complete model, 
but rather just assume that all masses
involved in our  fourth neutrino scheme are of the same order. 
However,  we would like to  
mention that there are indeed models where light sterile neutrinos do 
appear~\cite{pires,sterile}. 

The idea of using the sterile neutrino to improve the active neutrino mixings
was discussed earlier  in Ref.~\cite{balaji}, although in that paper the aim 
was rather to use a heavy sterile neutrino whose couplings to the active sector
could be tuned  to provide as much large mixing as possible, so one could start
with small mixings (as in the quark sector)  right  before considering the
sterile contributions.  
Here the idea we shall explore is quite different. 
We are suggesting  
that the sterile neutrino  could rather be
light, and yet have such couplings as to substantially
contribute to pull down an initially maximal solar mixing, and to generate
the mass splitting that provides the solar mass scale, particularly 
considering a class of models with the $L'$ symmetry. The idea may have a
realization for models with 
normal hierarchy too, but we are not exploring that in here.

To exemplify our idea, let us introduce a simple toy model for the sterile
couplings. 
We first consider the general form of the Majorana 
mass matrix in the basis ($\nu_\alpha,\nu_s$), 
 \be 
  M = \left( \begin{array}{c c}
           {\cal M}^{(0)} &  {\vec\delta}\,m\\
	{\vec\delta}\,^\dagger\, m & m_{s}
       \end{array} \right) ~,
  \label{mas}
 \ee
where $\vec \delta$ represent  
three parameters, smaller than unity,  that set the scale of
the active-sterile mass couplings. 
In general, these couplings would break $L'$, since the sterile carries no
standard lepton number. As we expect this couplings to play an important role
mainly in fixing the solar mixing, we will introduce the hypothesis that, after
bixamimal mixing, the sterile mainly couples to  one of the heaviest
states. Thus,  one can take for instance  
${\vec\delta}~^\dagger \sim 
(\delta/\sqrt{2}) (1,c_\theta,s_\theta)$. 
Some   small deviations of our choice will not
affect our conclusions. 
Moreover, to simplify our calculations even more we will assume that $m_s\ll
m$, so it can be neglected in the analysis. Any further 
model realization of the
mechanism we  present in here will have to provide an explanation for these
assumptions.  
Also, in last equation
${\cal M}^{(0)}$ stands for the active mass terms. Since we
continue assuming  they are generated in  models with the 
$L_e - L_\mu-L_\tau$ symmetry, it is natural to use  
the same form considered in the previous 
section. Therefore, at first order, we will again take  
${\cal M}^{(0)}= M_1$ as in Eq.~(\ref{m1}).

Within this approximations, and 
after performing a first rotation with the bimaximal mixing matrix, 
the four
by four mass matrix we have described gets the form
 \be 
 \tilde{M} =  
 m \left(\ba{cccc} 1 + \frac{\epsilon}{2} & -\frac{\epsilon}{2} & 0 & \delta\\  
 -\frac{\epsilon}{2} &-1 +\frac{\epsilon}{2}&0 &0\\ 0 &0&\epsilon&0\\
 \delta& 0&0&0 \ea\right)~.
 \label{mtilde}
 \ee
At zero order,  by taking  $\delta = \epsilon = 0$, it is clear that the 
mass matrix (\ref{mas}) is diagonal. 
The mass spectrum is of a 2+2 type, containing  
two heavy neutrinos, with masses $\pm m$ as before, and now two massless
neutrinos, one of them  the sterile neutrino which decouples in this limit. 
The gap among  both the sectors sets the atmospheric scale as usual.

 \subsection{Sterile contributions}

To better understand the role  the  sterile would play, 
let us first consider the 
special case where the breaking of the $L'$ 
symmetry is dominated by the sterile
contributions, that is when $|\epsilon|\ll |\delta|$. 
Thus, we will set $\epsilon=0$ for the moment. 
Now, after bimaximal rotation, 
the third state remains massless and decoupled 
from  the other neutrinos as well as  the second massive state. 
Hence, only the first and
fourth neutrino states remain mixed, with the mass terms
   \be 
  m~\left( \begin{array}{c c}
           1 &  \delta\\
     \delta & 0
       \end{array} \right) ~.
  \label{deltamas}
 \ee  
Now, the two eigenvalues of this matrix are 
$m_1 = \frac{m}{2} \left( 1 + \sqrt{1+4\, \delta^2}\right)
    \approx m(1+\delta^2)$ and 
$m_4 = \frac{m}{2} \left( 1 - \sqrt{1+4\, \delta^2}\right)\approx -m\delta^2$.
Therefore, at the lower order the sterile gets a see-saw type mass, setting a
new scale in the oscillation 
theory which is associated to $\Delta m^2_{34} = -m_4^2$. 
All other squared mass differences are 
$\Delta m^2_{13} \approx \Delta m^2_{14}\approx \Delta m^2_{23}
\approx \Delta m^2_{24} \approx \Delta m^2_{ATM}$; 
and for the solar scale we get 
 \be 
 \Delta m^2_{sol} = 
 \frac{1}{2}m^2 \left(  \sqrt{1+4\,\delta^2} -1 +2\,\delta^2 \right)
 \approx 2\,m^2\,\delta^2~,
 \ee
and thus, to get the right scale one needs
 \be 
 \delta^2 \approx \frac{1}{2}\frac{\Delta m^2_{sol}}{\Delta m^2_{ATM}}~. 
 \label{delta1}
 \ee
This equation implies that $\delta\sim 0.12$\,. 
Notice that above formula is similar to that in Eq.~(\ref{epsilon0}). 
This seems to indicate that  once  we include a  
non zero $\epsilon$, the solar scale should very
likely come due to the compensation or cancellation among  both the effects. 
The hierarchy $|\delta|>|\epsilon|$ is also suggested. 

On the other hand, the 1-4 mixing angle is  
 \be 
 \tan\beta = \frac{2\,\delta}{1 + \sqrt{1+4\, \delta^2}}\approx \delta~,
 \label{tanb}
 \ee
and therefore the total neutrino mixing matrix has the form
 \be 
 U_{mix} = U_{BM}\cdot U_{\beta} = 
 \left(\ba{cccc} 
  \frac{c_\beta}{\sqrt{2}}&\frac{1}{\sqrt{2}}&0&-\frac{s_\beta}{\sqrt{2}}\\
  \frac{c_\beta c_\theta}{\sqrt{2}} & -\frac{c_\theta}{\sqrt{2}}& s_\theta &
  -\frac{s_\beta c_\theta}{\sqrt{2}} \\
  \frac{c_\beta s_\theta}{\sqrt{2}}& -\frac{s_\theta}{\sqrt{2}}& -c_\theta &
  -\frac{s_\beta s_\theta}{\sqrt{2}}\\
 s_\beta & 0& 0& c_\beta   \ea \right) ~.
 \label{ubeta}
 \ee 
Notice that $U_{e3}$ remains zero due to the decoupling of the third massive
state. This also cancels any contribution of the new scale $\Delta m^2_{34}$
into the electron neutrino survival probability, $P_{ee}$, 
since the corresponding mixing angle would be $4 (U_{e3}U_{e4})^2 =0$.
There are, nevertheless,  electron sterile neutrino oscillations only via the  
atmospheric scale. Indeed from above mixings one gets 
$P_{es} =  \frac{1}{2} \sin^2 2\beta \sin^2(\Delta m^2_{ATM} L/4E)$. 
However, the corresponding mixing angle is small at this limit,  
of order $\delta^2\sim10^{-2}$,  so we will not discuss it further.

Solar mixing angle now becomes
 \be 
 \sin^2 2\theta_{sol} = \cos^2\beta = 
 \frac{1 + \sqrt{1+4\, \delta^2}}{2 \sqrt{1+4\, \delta^2}} 
 \approx 1 - \delta^2 ~.     
 \label{sinsterile}
 \ee
From here we can write  $\sin 2\theta_{sol}= \sin(\frac{\pi}{2} -\beta)$.
Hence 
\[\theta_{sol} = \frac{\pi}{4} - \frac{\beta}{2}~,\] 
and 
so we notice the  sterile corrections  also work in the right direction,
they reduce the solar mixing from its maximal value. However, as before,
one can see that the effect is not yet enough to pull
$\theta_{sol}$ down to the desired values. Indeed for central values of the 
squared mass differences one gets
$\sin^2 2\theta_{sol} =  0.984$, which corresponds to 
$\sin^2 \theta_{sol} = 0.437$. 

\subsection{Active-Sterile compensation: The fall of the solar mixing}

From previous sections  one sees 
that neither the  active nor the sterile corrections we
assumed are capable enough by themselves to provide the 
right corrections for both the solar parameters. 
Therefore, we will now consider both scenarios together to show how the
correct values may arise when both mechanisms are at work.
Next we reinsert the $\epsilon$ parameter on $\tilde M$.
Motivated by the suggestive hierarchy $|\delta|>|\epsilon|$, 
discussed in  previous section, we will proceed as follows. 
First we  write
 \be 
 {\tilde M}  = {\tilde M}_0 + {\tilde M}_\delta + {\tilde M}_\epsilon~,
 \ee
where ${\tilde M}_{\delta,\epsilon}$ only  contains the $\delta$ and $\epsilon$
contributions respectively. ${\tilde M}_0$ is our diagonal 
zero order mass matrix, as before.  Next we change our basis into that
where ${\tilde M}_0 + {\tilde M}_\delta$ is diagonal, 
and treat 
${\tilde M}_\epsilon$ as a perturbation. 
Clearly this amounts to simply rotate $\tilde M$ on
Eq. (\ref{mtilde}) by the same 1-4 mixing  we calculated 
in previous section [Eq.~(\ref{ubeta})], without fixing $\delta$ yet. 
We get
$U_\beta^\dagger\cdot(\tilde{M}_0 + {\tilde M}_\delta)\cdot U_\beta 
  = m\cdot{\rm diag}\{\lambda_+\, ,\, 
  -1\,,\,0\,,\, \lambda_-\}$; for 
$\lambda_\pm = \frac{1}{2}(1\pm\sqrt{1 + 4\, \delta^2})$;  
and for the perturbation matrix:
 \be 
  U_\beta^\dagger\cdot\tilde{M}_\epsilon\cdot U_\beta =  
 m\, \frac{\epsilon}{2} 
 \left(\ba{cccc} c_\beta^2 & -c_\beta & 0 & -s_\beta\,c_\beta\\  
 -c_\beta &1&0 & s_\beta\\   0 &0 &2&0\\  
 -s_\beta\,c_\beta& s_\beta&0&s_\beta^2 \ea\right)~.
 \label{mepsilon}
 \ee
Of course, in the limit $\beta=0$ ($\delta=0$) 
last expression becomes Eq.~(\ref{me}), indicating that at 
the  leading order 
the effect is effectively to rotate the 1-2 sector.
Also, it is worth noticing that the third state always decouples, which means
that $U_{e3}=0$. Thus, although there is a new scale in our final model, 
$\Delta m^2_{34}$, it would not contribute to $P_{ee}$ either. 

The first order mass eigenvalues one gets are
 \bea
 m_1 &\approx& \frac{m}{2} \left( 1 + \sqrt{1+4\, \delta^2}\right) +
 m\,\frac{\epsilon}{2}\,\cos^2\beta~; \\ 
 m_2 &\approx& m\left(-1 + \frac{\epsilon}{2}\right)~;\\
 m_3 &=& \epsilon\, m~;\\
 m_4 &\approx& \frac{m}{2} \left( 1 - \sqrt{1+4\, \delta^2}\right) +
 m\,\frac{\epsilon}{2}\,\sin^2\beta~.
 \eea
{}From here, the splitting among the first two states is
 \be 
 \Delta m^2_{12} \approx \left[ \delta^2 - \lambda_- + 
 \epsilon\left(1+\lambda_+~\cos^2\beta\right)\right]m^2~.
 \label{solardm}
 \ee
We can approximate last equation at first order in both parameters  to get the 
more simple expression
 \be 
  2\delta^2 + \epsilon\left(1 + \cos^2\beta\right)\approx 
 \frac{\Delta m^2_{sol}}{\Delta m^2_{ATM}}~;
 \ee
from where it is easy to see that for a relatively large $\delta$ 
(say of order 0.3),
$\epsilon$ has to be negative in order to compensate the effect. 
$\delta$ can well be positive or negative, with out any modification on the
results. We will assume it positive. This way solar
scale would arise as due to the cancellation among both corrections. On the
other hand, this would allow the solar mixing to get larger corrections from 
$\delta^2$, as it is suggested from Eqs.~(\ref{solmixa1}) and 
(\ref{sinsterile})
since quadratic $\delta$ corrections are expected to dominate. 

It is worth noticing that for other mass squared differences we get 
$\Delta m^2_{13}\approx \Delta m^2_{14} \approx\Delta m^2_{23}\approx
\Delta m^2_{24}\approx \Delta m^2_{ATM}$; and 
\be
|\Delta m^2_{34}|\approx 
\left|\epsilon^2 -
\left(\lambda_- +\frac{\epsilon}{2}\sin^2\beta\right)^2\right|m^2
\ll \Delta m^2_{sol}~.
\ee

Next, final mixing is given by  
$U_{mix} = U_{BM}\cdot U_\beta\cdot U_{\epsilon}$, where $U_\epsilon$ is the
extra rotation needed to completely  diagonalize $\tilde M$ in 
Eq.~(\ref{mtilde}), which at first order in $\epsilon$ is given by
 \be 
  U_\epsilon\approx   
 \left(\ba{cccc} 
 1 & \frac{c_\beta\epsilon}{2(1+\lambda_+)}& 
             0 & \frac{c_\beta s_\beta\epsilon}{2(\lambda_+ - \lambda_-)}\\  
 -\frac{c_\beta\epsilon}{2(\lambda_+ +1)}&1&
             0 &\frac{s_\beta\epsilon}{2(1+\lambda_-)} \\   
 0 &0 &1&0\\  
 -\frac{c_\beta s_\beta\epsilon}{2(\lambda_+ - \lambda_-)} & 
 -\frac{s_\beta\epsilon}{2(1+\lambda_-)}&0&1 \ea\right)~.
 \label{ue}
 \ee
{}From here, the solar mixing is calculated to first order and gives
 \be 
 \sin^22\theta_{sol} \approx
 \cos^2\beta\left[1 + \frac{3}{2}\epsilon\sin^2\beta
    \left(\frac{1+\delta^2}{(2- \delta^2)\sqrt{1+4\delta^2}} \right)\right]^2~.
 \label{solarmix}
 \ee
Leading correction is of order $\delta^2$, as expected. 

Unlike the previous results, where independent effects were studied, now the
contribution of both, active $L'$ breaking effects and sterile couplings give us
enough freedom to account for both the  solar parameters.
We use Eqs.~(\ref{solardm}) and (\ref{solarmix}) 
to get a rough estimation of the required values for our parameters 
in order to get central values for solar scale, 
$\Delta m^2_{12} = 7.92\times 10^{-5}~{\rm eV^2}$;
and  solar
mixing, $\sin^2\theta_{sol}  = 0.314$ 
and get $\delta  = 0.412$ and  $\epsilon=-0.141$, which nicely
validate our approximations. 
Moreover, with this values we get for  the new
scale  about $|\Delta m^2_{34}| = 9.5\times 10^{-6}~eV^2$, which is a factor
of eight smaller than the solar scale.

\begin{figure}
\centerline{
\epsfxsize=200pt
\epsfbox{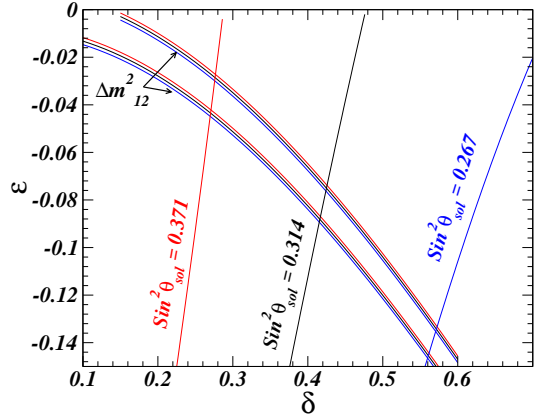}
}
\vskip1ex

\caption{Allowed values of $\epsilon$ and $\delta$ which give the solar
parameters, $\Delta m^2_{12}$ and $\sin^2\theta_{sol}$;
within two sigma deviations respect to observed central values. The two narrow 
regions for the squared mass difference are depicted.}

\end{figure}

A more precise calculation 
can be done starting from the mass matrix in Eq.~(\ref{mtilde}), by performing 
a numerical diagonalization for given values of the pair $(\delta, \epsilon)$.
This allows in general to identify the range of parameters 
that would give acceptable  results for our model. We present our results in
Fig.~1, where we plot the 
bands for which solar scale and mixing are obtained within two 
sigma deviations. 
Precise results substantially reduce the required value for
$\epsilon$, as it can be seen from the plot.
Notice that whereas the mixing angle, $\sin^2\theta_{sol}$, is little
sensitive to $\epsilon$, and corresponds to a wide region on the
parameter space, there are two
allowed regions for the solar scale, $\Delta m^2_{12}$,
that are very narrow and clearly 
sensitive to both $\epsilon$ and $\delta$ parameters. 
This indicates that some level of fine tuning 
may be required to get the proper scale.

\section{Confronting the sterile with experimental data}

Most studies we are aware of 
on the constrains for sterile neutrinos are based in the
hypothesis that these are heavier than any active scale (see for instance
Ref/~\cite{vissani}). However, 
sterile mixing to electron neutrino at a scale somewhat
smaller than solar scale was consider in Ref.~\cite{holanda}, and there it was
shown that a
weak mixing is yet allowed.
An extended and detailed
study for light sterile neutrinos 
is still required and will be presented in a 
forthcoming paper. 
Here we will only make some remarks on what the potential constrains would be,
based on the simple model we have presented, and simply extrapolating the
results of the somehow general analysis given in Ref.~\cite{vissani}. 

First of all, the appearance of a new light scale, 
$\Delta m^2_{34}$, of order $10^{-5}~{eV^2}$
may suggest a new level crossing in the resonant conversion of electron
neutrinos within the Sun. However, by looking at the general 
survival probability for electron neutrinos (assuming no CP violation),
 \be
 P_{ee} = 
 1 - 4\sum_{a>b}^4\sum_{b=1}^4 
 \left( U_{ea}  U _{eb}\right)^2 
 \sin^2\left(\frac{\Delta m_{ab}^2 L}{4E}\right)~,
 \ee
where we have used $U$ in place of $U_{mix}$ for simplicity; 
it is easy to see that the mixing angle associated to  
$\Delta m^2_{34}$ is 
$4\left[ U_{e3} U_{e4}\right]^2=0$, 
since $U_{e3}$ is exactly zero in above results.
Therefore, as we already mentioned, 
only solar and atmospheric scales would contribute to the
oscillations. This does not mean, however, the absence of electron to sterile
conversion, which indeed appears as indicated by
the oscillation probability formula
 \be 
 P_{es} = 4\sum_{a>b}^4\sum_{b=1}^4 
 U_{ea}  U_{eb} U_{sa} U_{sb}
 \sin^2\left(\frac{\Delta m_{ab}^2 L}{4E}\right)~.
 \ee
From here, it is again clear that since
$U_{e3}=U_{s3}= 0$, 
there is no contribution from the 
$\Delta m^2_{34}$ scale. Resonant conversion may only occur at solar scale,
however, the associated mixing 
 \be
 4U_{e1} U_{e2}U_{s1}  U_{s2}
 \approx -\frac{1}{2}\sin2\theta_{sol}\, c_\beta\, s_\beta^2\epsilon 
 \approx 10^{-2}~,
 \ee 
corresponds to a point on parameter space, with $\tan^2\theta_s\approx 10^{-3}$,
that is not at all excluded, but
marginally accepted, as it can be seen from the analysis presented in
 Ref.~\cite{vissani}, if we naively extrapolate their results. 
Point is that the dip of the survival electron 
probability produced for the resonance is dominated by the mixing to active
neutrinos.
We believe, nevertheless, that a detailed study for our case is
required to be conclusive.
Just to  get a naive idea of the effect,  
we can roughly estimate the fraction of sterile neutrinos in the solar flux
using an ``in vacuum'' approximation assuming $4E/L$ at  the solar scale, 
to get
 \be 
  \eta_s = \frac{P_{es}}{1-P_{ee}}\approx  7.1\times 10^{-2}~;
 \ee
which is just below current limits (see for instance Ref.~\cite{vissani}).
Although somehow unjustified due to the matter effects, 
we believe this calculation should give a close result 
to the correct one, and provides an  indication of the marginal acceptance of
our model.

Non resonant conversion will also occur associated to atmospheric scale, 
but this contribution is expected to be less relevant 
at solar neutrino energies. 

The analysis done by Cirelli {\it et al.} 
in Ref.~\cite{vissani} also serves very well for a first check of
consistency of our model with Supernova and cosmology constrains, by taking a
naive extrapolation of their results to our case.
Supernova (SN) produces neutrinos and anti-neutrinos in roughly 
a similar amount, however, present experiments focus mainly in anti-neutrinos,
for which there is no resonance matter effects. This is reflected in almost no
restrictions for the  above small $\nu_e-\nu_s$ mixing, also  
relevant at SN neutrino energies, as it can be checked in the mentioned
reference.
Main cosmological bounds on 
sterile neutrinos come from Big Bang Nucleosynthesis, which probes the total
energy density at $T\sim 0.1 - 1$~MeV, constraining the number of relativistic
species at that energy, $N_\nu$ (for a review see~\cite{sarkar}). A larger
effective number contribution of $N_\nu$ than three would 
affect  $^4He$ and Deuterium abundance, in a rather mild way. Conservative
estimates give $N^{^4He}_\nu \approx 2.4\pm 0.7$ and $N^{D}_\nu \approx 3\pm 2$.
Currently CMB does not give strong bounds~\cite{cmb,steen}: 
$N_\nu\approx 3\pm 2$. Finally, structure formation imposes a constrain on the
total energy density in neutrinos, $\Omega_\nu h^2$, from where cosmological
bound to absolute neutrino mass is obtained.
However, as one can check from Cirelli {\it et al.}
results,  light sterile neutrinos with small mixings
contribute little to $N_\nu$. In fact one gets $N_\nu$ just around 3.2.  
Also, no important bounds come from structure
formation either due to the small sterile mass.

\section{Concluding remarks}

We have suggested the possibility that light sterile neutrinos 
may be the missing ingredient that transforms a maximal mixing 
into the observed large mixing angle in solar neutrino oscillations.
To elaborate the idea 
we have presented a simple toy model that realizes this possibility with a
sizable effect that may accommodate the observed values, in models with inverted
hierarchy provided by a broken global $L_e -L_\mu - L_\tau$ symmetry.
Models with this symmetry usually face the problem of predicting a too large 
solar mixing angle. With the 
introduction of appropriate sterile couplings the models gain enough freedom to 
arrange both solar 
parameters to the proper order of magnitude.
We believe our mechanism may also work for models with normal hierarchy which
we have not explored in here, though.

The light sterile neutrino should come with  
a new mass scale, $\Delta m^2_{34}$, which in order 
to produce the desired corrections
comes out  to be lighter than solar scale.
However, in the toy model we have explored the new scale
does not contribute to 
electron neutrino survival probability in solar oscillations, since it is
attached to $U_{e3}$ which is exactly zero. Thus, no resonant
conversion of solar neutrinos is expected for these scale. 
Moreover, the effective
fraction of sterile neutrinos in the solar flux seems to be just below the
current limits. Indeed, a very rough  
calculation using central experimental
 values and vacuum oscillations 
gives the prediction $\eta_s\approx 7.1\times 10^{-2}$.
Cosmology and Supernova bounds seem to marginally alow the light sterile
neutrino
with the parameters we have used.  
A cautionary word should be given. We are not presenting a complete model yet,
but rather exploring just the basics of the  idea. 
A more detailed analysis of atmospheric and solar neutrino oscillations, 
including matter effects is desirable
to precise the constraints
and determine whether future experiments would be
sensible to the new scale. Those results may 
even  rule out the present scenario, but perhaps may suggest a more realistic
one. 
A light sterile neutrino seems to be  anyway  an interesting possibility that
deserves further exploration.  


\section*{Acknowledgments}
APL would like to thank the warm hospitality of 
The Abdus Salam ICTP, at Trieste, Italy, 
where part of this work was done.
We also thank A.Yu. Smirnov for comments on the early stages of the 
work and also to G. O. Miranda. 
This work has been  
partially supported by CONACyT, M\'exico, under grant J44596-F.


\end{document}